\documentclass{webofc}
\usepackage[varg]{txfonts}   
\usepackage{amsmath}
\usepackage{amssymb}
\usepackage{caption}
\usepackage{subcaption}
\usepackage{xcolor}
\usepackage{graphicx}
\graphicspath{ {./} }
\newcommand{\comment}[1]{}

\begin{document}
\title{What Machine Learning Can Do for Focusing Aerogel Detectors}
%
%

\author{\firstname{Foma} \lastname{Shipilov}\inst{1}\fnsep\thanks{\email{foma@shipilov.ru}} \and
        \firstname{Alexander} \lastname{Barnyakov}\inst{2,3} \and
        \firstname{Viktor} \lastname{Bobrovnikov}\inst{2} \and
        \firstname{Sergey} \lastname{Kononov}\inst{2,4} \and
        \firstname{Fedor} \lastname{Ratnikov}\inst{1}
}

\institute{NRU Higher School of Economics, Moscow, Russia
\and
           Budker Institute of Nuclear Physics of Siberian Branch Russian Academy of Sciences, Novosibirsk, Russia
\and
           Novosibirsk State Technical University, Novosibirsk, Russia
\and
           Novosibirsk State University, Novosibirsk, Russia
          }

\abstract{%
Particle identification at the Super Charm-Tau factory experiment will be provided by a Focusing Aerogel
Ring Imaging CHerenkov detector (FARICH). The specifics of detector location make proper cooling difficult, therefore a significant number of ambient background hits are captured. They must be mitigated to reduce the data flow and improve particle velocity resolution. In this work we present several approaches to filtering signal hits, inspired by machine learning techniques from computer vision.
}
\maketitle
\section{Introduction}
\label{intro}

\begin{figure}[h]
    \centering
    \includegraphics[width=0.55\linewidth]{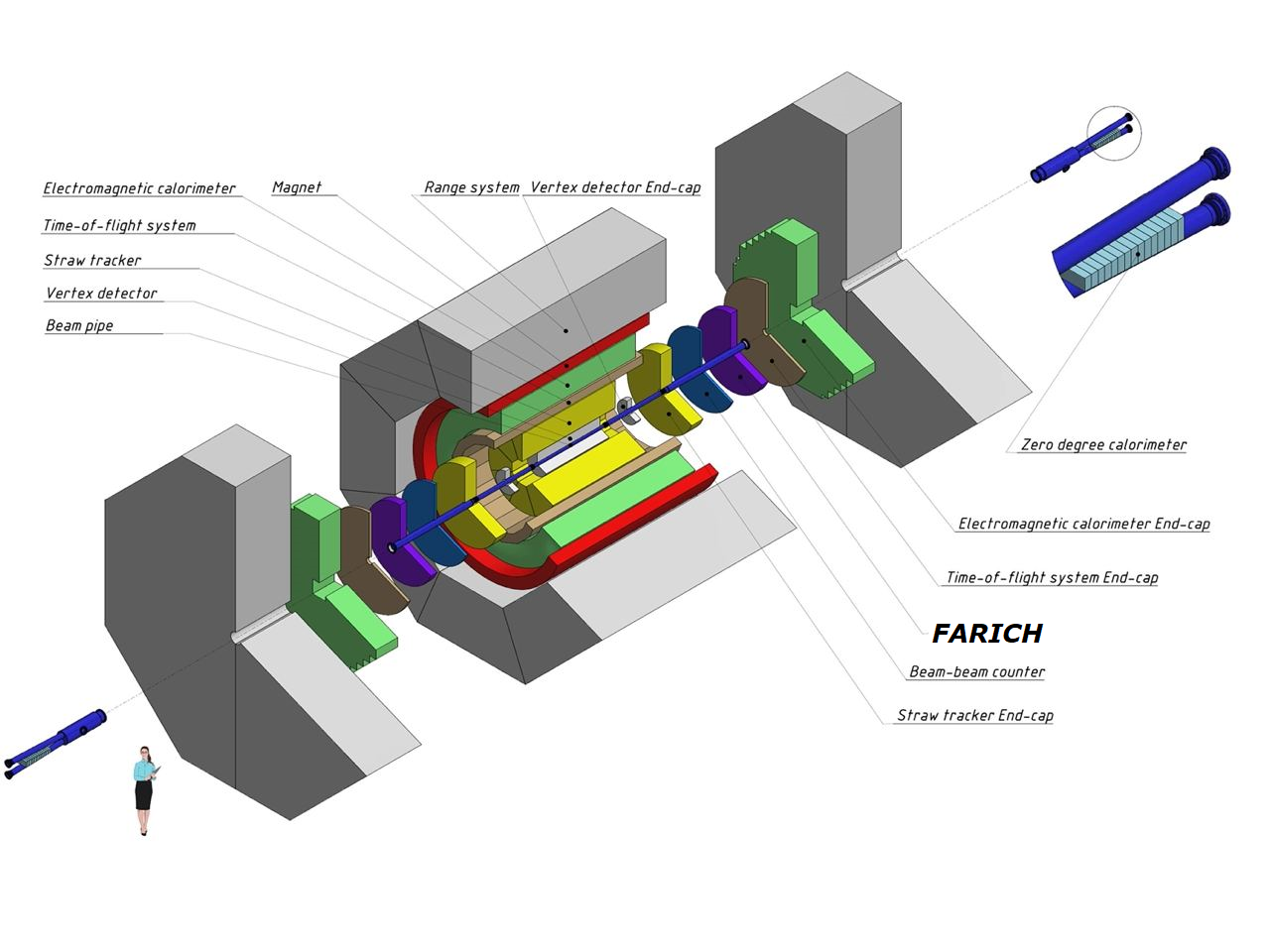} \quad
    \raisebox{0.3\height}{\includegraphics[width=0.3\linewidth]{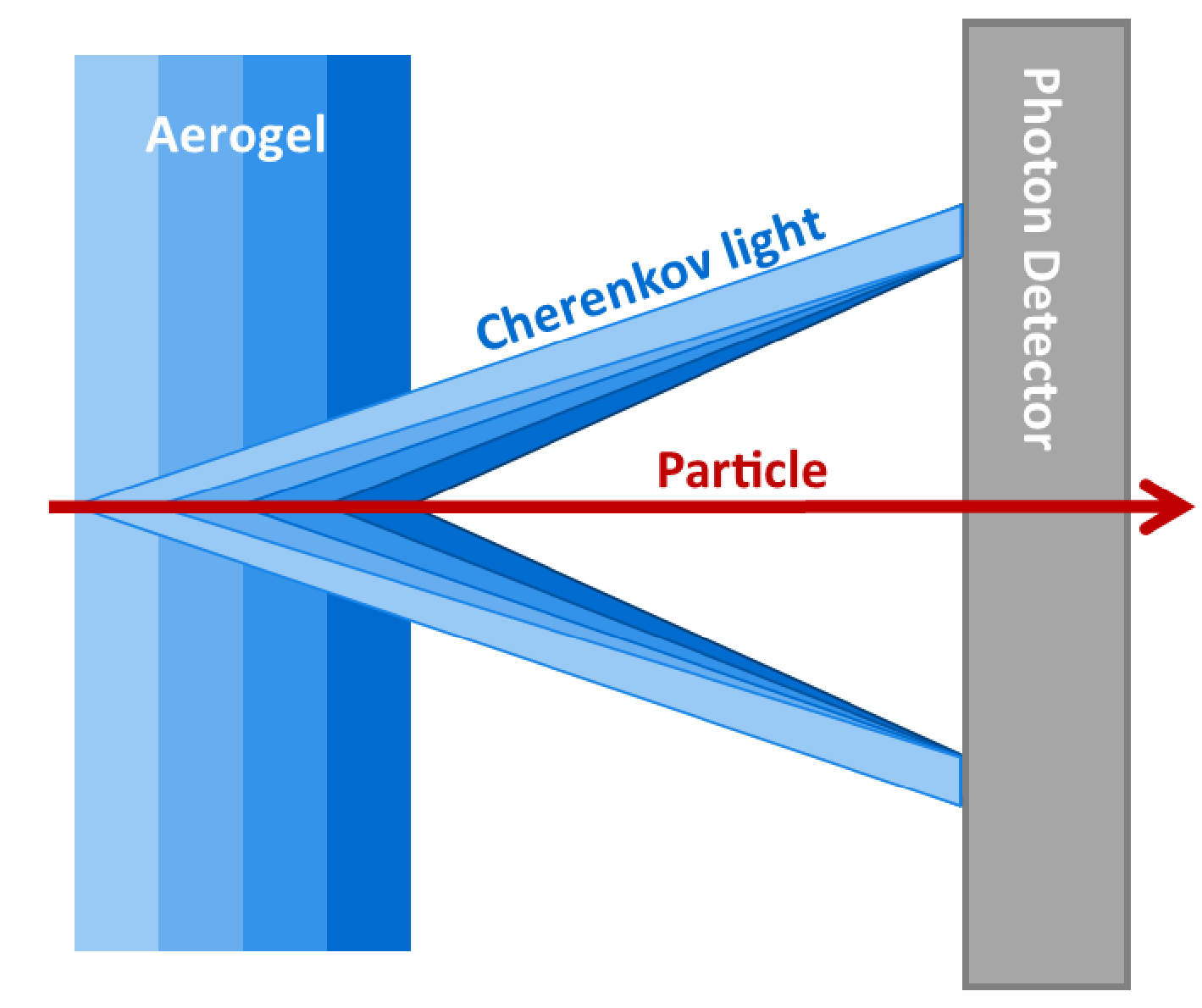}}
    \caption{Left to right: possible integration of FARICH in SPD detector \cite{korzenev2023spin}, focusing aerogel operation \cite{Barnyakov_2020}.}
    \label{fig:farich}
\end{figure}

Reliable particle identification (PID) is a crucial component of modern physics experiments. Particle identification in the Super \textit{c}-$\tau$ factory (SCTF) experiments will be provided by FARICH detector \cite{2018PPN....49...30F}. The use of a FARICH is under intensive discussion for the Spin Physics Detector (SPD) detector at NICA \cite{korzenev2023spin}. FARICH uses multilayer aerogel for Cherenkov ring proximity focusing (Fig. \ref{fig:farich}). The detector may use both seedless real-time signal finder to produce fast trigger and mitigate noise background, and seeded off-line reconstruction mode for precise identification, however, SiPM properties and operating temperatures may result in a significant background hit rate $f \sim$1 MHz/mm$^2$ (signal-to-background ratio $\approx0.014$), which necessitates the development of robust noise filtering techniques.

Conventional pattern recognition noise rejection methods aim to eliminate background hits directly by calculating carefully crafted empirical statistics derived from the physical properties of the system \cite{drielsma2021scalable}, e.g. DELPHI RICH automates background removal by measuring photon hit times, the number of hits for each detector cell, and other parameters \cite{BLOCH1996236}.

Statistical approaches to pattern recognition and reconstruction provide excellent precision and enable simple error estimation. However, such methods are only effective in moderate background conditions \cite{BLOCH1996236} due to the local nature of the statistics. This does not suit well for the heavy background environment of the FARICH. Additionally, statistical approaches require a particle track prior and may be difficult to incorporate new data sources into.

In contrast, Machine Learning (ML) techniques can extract high-level features from the input data and use them to efficiently solve various tasks. Many different data sources can be relatively easily incorporated, contributing to the overall robustness of the method.

ML approaches proved to be useful across a diversity of nuclear physics research topics \cite{Boehnlein_2022}. Early variants of neural networks had been developed specifically for use in high energy physics \cite{lonnblad1992pattern}. LHCb experiment uses neural networks for fast fake track rejection \cite{albertsson2018machine} and is planning to incorporate them in the RICH PID system \cite{blago2023deep}. ML has been applied to calibration and reconstruction of Cherenkov detectors with promising results \cite{Fanelli_2020}. Recently, object detection techniques from computer vision have been utilized in end-to-end data reconstruction pipeline for LArTPC neutrino imaging \cite{drielsma2021scalable}. Object detection has also been adapted for sparse detector data in the object condensation pipeline \cite{kieseler2020object}.

In this work, we present our ML-based approach to noise filtering and event reconstruction of the FARICH detector. \comment{For the filtering task, we use a Convolutional Neural Network (CNN) to perform binary classification.}

\section{Data processing and model architecture}
\label{sec-data}

\begin{figure}[h]
    \centering
    \includegraphics[height=0.41\linewidth]{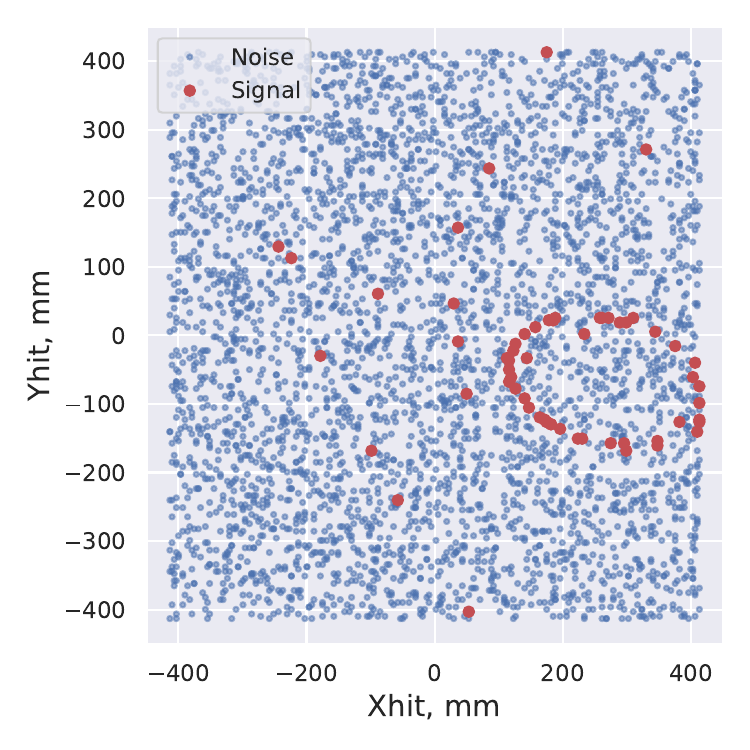}
    \includegraphics[height=0.41\linewidth]{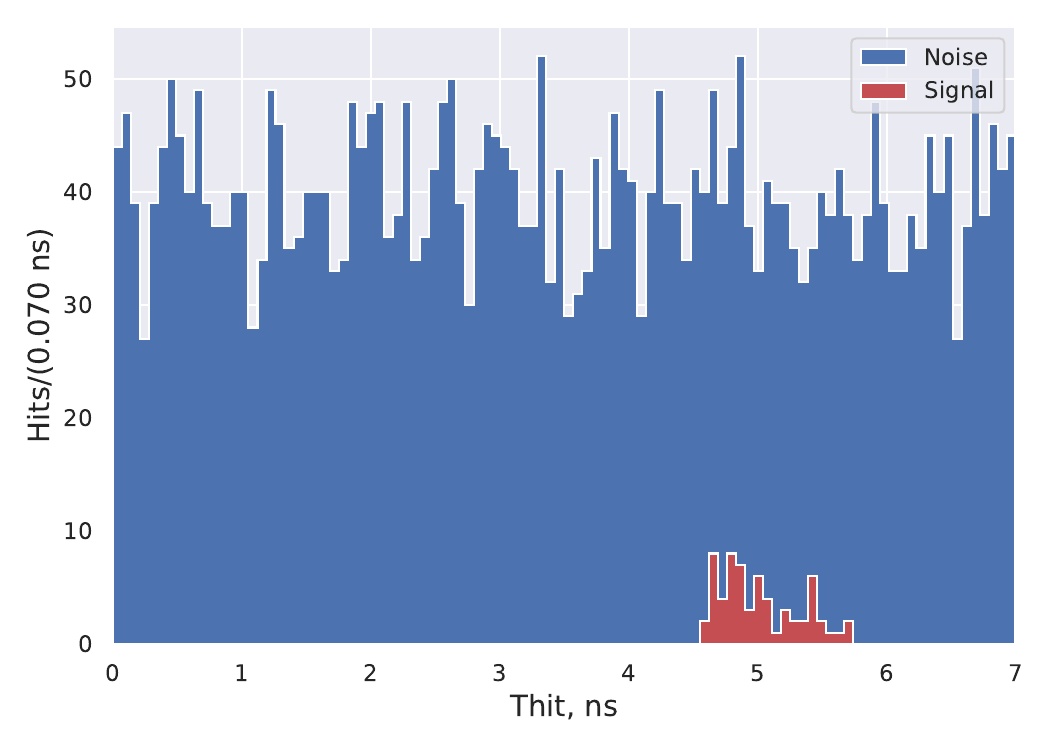}
    \caption{Event example for $f=1$MHz/mm$^2$. The field is $240\times240$ pixels wide; event duration is $7\cdot10^{-9}$ s.}
    \label{fig:data}
\end{figure}

The signal data is generated in a Geant4 \cite{allison2006geant4,agostinelli2003geant4} simulation (Fig. \ref{fig:data}). The data contains the coordinates $x_c,y_c$ of detected photons in a SiPM grid and hit times $t_c$. We utilize the ability of ML to handle different data sources in a unified fashion to feed the model both coordinates and times in a 2-channel image format, where the first channel is a bilinearly interpolated binary mask computed from $x_c,y_c$ and the second channel consists of normalized $t_c$ for the corresponding pixels. Signal hits tend to form a compact cluster in the time dimension ($\sim2$ ns compared with $\sim7$ ns for the whole event), which can aid the search for a signal pattern. Uniform random noise is then applied on top of the signal to simulate ambient background hits.

We use ResNet-18 \cite{https://doi.org/10.48550/arxiv.1512.03385} and make changes to the input and output layers to accommodate for our data formats.

\section{Noise filtering}
\label{sec-filter}

The noise filtering task can be formulated in terms of binary classification of events with / without signal, which allows us to use a simple logistic regression objective. Ground truth classes are determined as follows: first, a bounding box enclosing the signal ellipse is computed from particle track parameters, then, events are labeled positive if they contain $\geqslant10$ signal photons. These parameters result in a class balance ratio of $3 / 7$ (positive / negative).

The metrics of our primary concern are efficiency and noise events reduction, which can be mapped nicely to true positive rate (the ratio of true positive to all positive samples) and false positive rate (the ratio of false positive to all negative samples) correspondingly in a classification task.

Thanks to this fact, we can analyze performance of the model by comparing standard receiver operating characteristic (ROC) curves. Area under curve (AUC) can be used as a validation metric in model selection (hyperparameter search).

We also train weighted binary classification with various positive class weight values (Eq. \ref{eq:logreg}). Class weight is usually introduced to counter class imbalance, instead, we try to control the trade off between our primary metrics, with the motivation that aiming for specific performance during training might potentially lead to better results compared to varying classifier threshold after the training.
\begin{equation}
    \label{eq:logreg}
    \mathcal{L}(x, y)=-w_{\text{pos}}y\log(\theta(x)) - (1 - w_{\text{pos}})(1 - y)\log(1 - \theta(x)),
\end{equation}
where $x$ stands for input, $y$ -- target, $\theta$ -- probability estimator (CNN), $w_{\text{pos}}$ -- positive class weight.

The CNN is initialized with pretrained weights from the reconstruction task to speed up learning.

\subsection{Experiment}
\label{subsec-filter-exp}

We train several models with a diverse set of $w_{\text{pos}}$ and $f\in[100\text{ KHz}; 1\text{ MHz}]$.

\begin{figure}[h]
    \centering
    \begin{subfigure}[b]{0.5\linewidth}
        \includegraphics[width=1\linewidth]{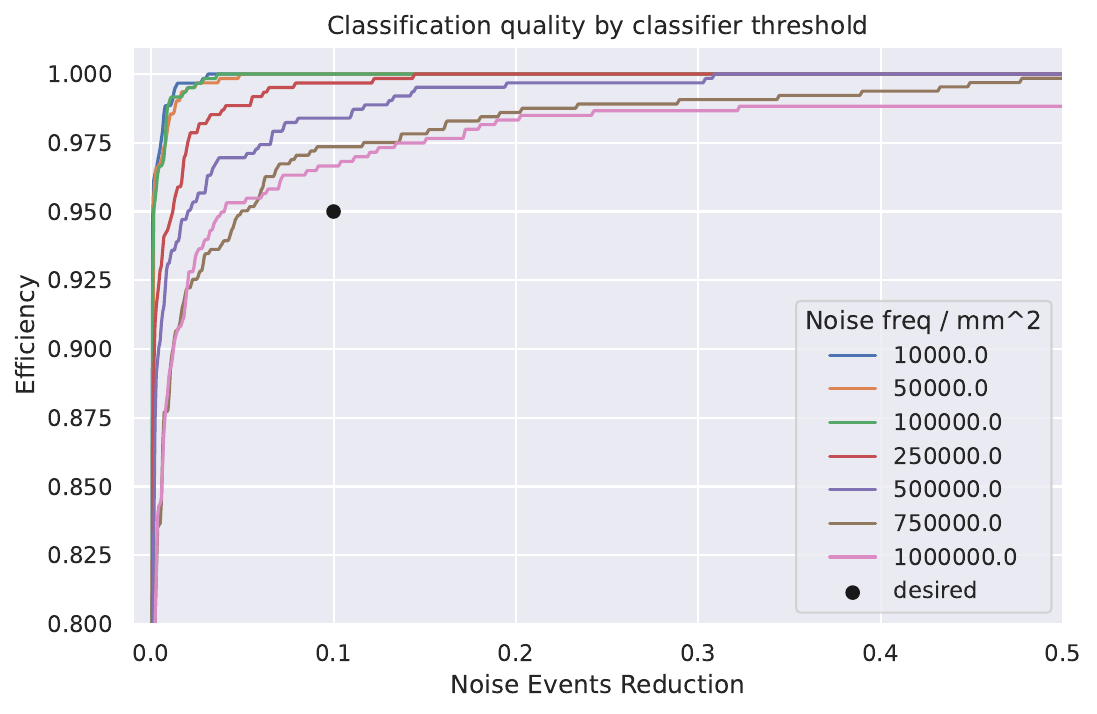}
        \caption{Classification: ROC curves for different $f$ values.}
        \label{fig:posthr_a}
    \end{subfigure}
    \hfill
    \begin{subfigure}[b]{0.4925\linewidth}
        \includegraphics[width=1\linewidth]{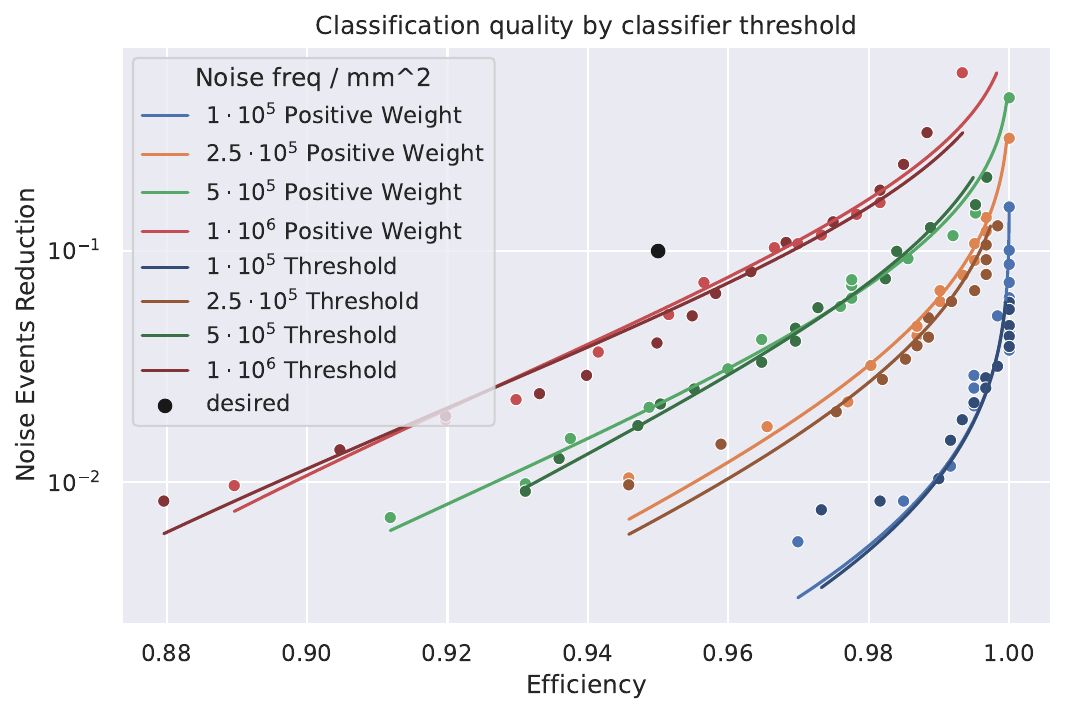}
        \caption{Bounding box regression: Comparison of $w_{\text{pos}}$ and threshold setting.}
        \label{fig:posthr_b}
    \end{subfigure}
    \caption{Validation noise filtering performance.}
    \label{fig:posthr}
\end{figure}

We demonstrate that our method provides a significant level of noise rejection with high efficiency (Fig. \ref{fig:posthr_a}). For $\pi^{-}$ events, $f = 10^6\ \text{Hz}/\text{mm}^2$ 90\% background rejection has been achieved, whilst 95\% of the signal has been retained. No significant difference in performance was observed between training with specific $w_{\text{pos}}$ and setting the threshold in the default logistic regression (Fig. \ref{fig:posthr_b}). Therefore, it is preferable to use the simpler method and avoid time-consuming training of multiple models.

The model performs the best for high momentum events (Fig. \ref{fig:posthr_a}). The quality significantly decreases on the lower end of momentum spectrum, as well as for angles of incidence closer to tangent.

To integrate a CNN model into the filtering pipeline, one can restate the problem in terms of bounding box regression, that is, predicting the coordinates of a rectangular box bounding the signal \cite{shipilov2023machine}. In comparison to this approach, classification is less precise, but much more robust (Fig. \ref{fig:posthr}). It is possible to achieve even higher noise reduction by chaining bounding box regression with classification for positive events.

\begin{figure}[h]
    \centering
    \includegraphics[width=1\linewidth]{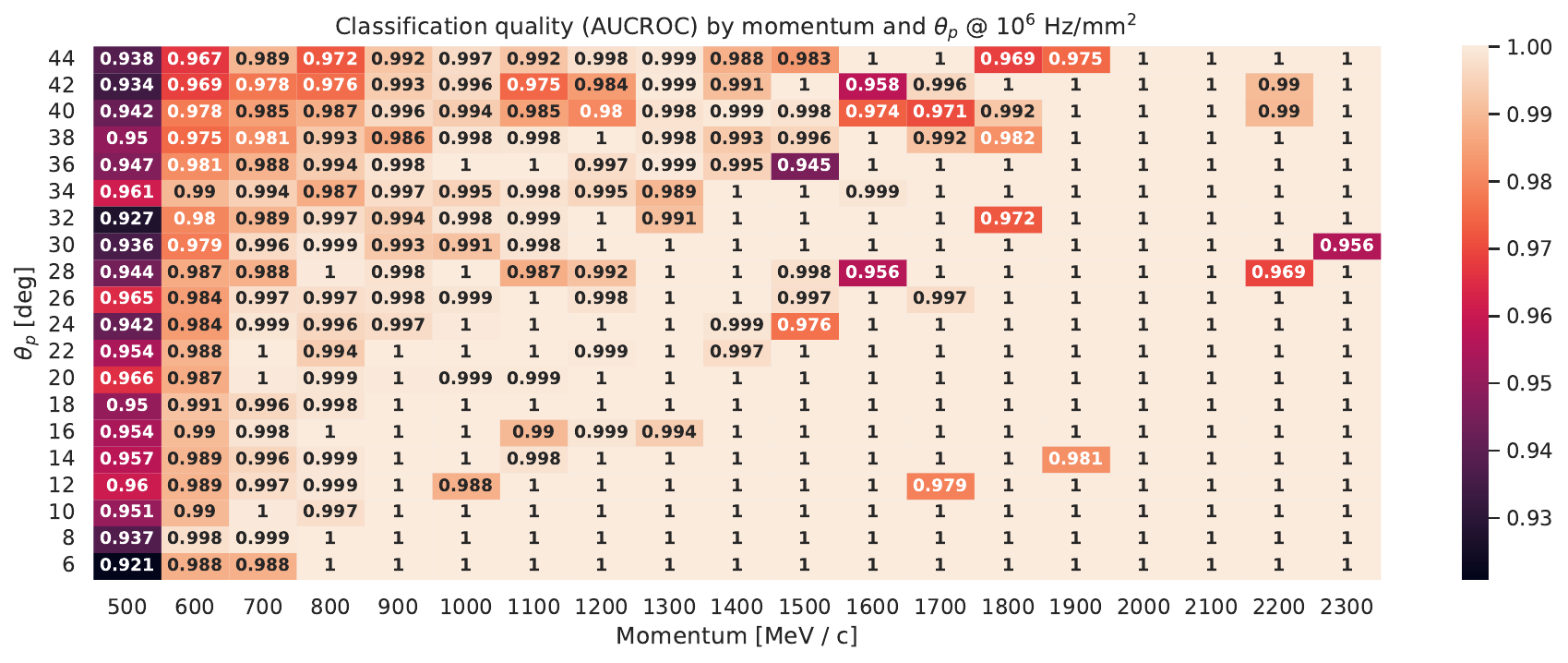}
    \caption{Classification performance by momentum $p$ and angle of incidense $\theta_p$.}
    \label{fig:mom_theta_p}
\end{figure}


\section{Summary}
\label{sec-sum}



The process of developing the FARICH detector gives rise to a range of essential problems: fast online noise filtering, offline reconstruction, fast simulation, etc. Machine Learning (ML) proved to be helpful across a wide range of applications in high energy physics, particularly for RICH detectors, hence, applying ML for FARICH has a great potential. Moreover, FARICH may be adapted for the second phase of NICA SPD experiment, which further expands the applicability of our research. The initial results of applying ML for FARICH noise filtering are promising, with a lot of useful implications and potential for future studies.

\begin{acknowledgement}
The research leading to these results has received funding from the Basic Research Program at the National Research University Higher School of Economics.
\end{acknowledgement}
%
\bibliography{template.bib}
%
%
\comment{}

\end{document}